\documentclass{ws-procs9x6-cpt16}

\begin{document}

\newcommand{\refeq}[1]{(\ref{#1})}
\def\etal {{\it et al.}}
%any other macros go here 

\def\sla#1{\hbox{{$#1$}\llap{$/$}}}

\title{Lorentz-Invariance Violation with Higher-Order Operators}

\author{Carlos M.\ Reyes$^1$ and Luis F.\ Urrutia$^2$}

\address{$^1$Departamento de Ciencias B{\'a}sicas, 
Universidad del B{\'i}o-B{\'i}o\\ 
Chill\'an, Casilla 447, Chile }
\address{$^2$Instituto de Ciencias Nucleares, 
Universidad Nacional Aut{\'o}noma de M{\'e}xico\\
Ciudad de M{\'e}xico\ 04510, M{\'e}xico. }

\begin{abstract}
In this work, in the light of the developments 
for indefinite metric theories made by Lee and Wick, we study
perturbative unitarity in a Lorentz-invariance violating
QED model with higher-order operators. We show that by following the Lee-Wick prescription
 it is possible to preserve unitarity in the model at one-loop order in the coupling.
\end{abstract}

\bodymatter

\section{Introduction}
A direct generalization 
of the Standard-Model Extension\cite{KS} (SME) follows by  
taking into account nonrenormalizable operators, that is, operators with mass dimension 
higher than four. 
Such program has been successfully implemented in the 
photon sector,\cite{KM1} fermion sector\cite{KM2} and more recently in the linearized sector 
of gravity.\cite{KM3}
An earlier work of Myers and Pospelov focuses
 on dimension-five operators with approximately cubic dispersion relations.\cite{MP}
Alternatively, Lorentz-invariance violation with higher-order operators
may be realized with
higher-order coupling terms.\cite{MC}

Quantum field theory with higher-order operators may lead
to an indefinite metric in the Hilbert space.
The extended inner product introduced by the 
indefinite metric $\eta$ allows for negative norm states
or ghosts and produces
a pseudo-unitary condition for the $S$ matrix, i.e.,
$S^{\dag}\eta S=\eta$. 
As shown by Lee and Wick, by
imposing the boundary condition 
in which only positive norm states appear in the asymptotic Hilbert space, it is possible to preserve unitarity order by order
in perturbation theory.\cite{LW}
In this work, in the light of the Lee-Wick studies, we show how unitarity can be conserved in 
 a nonminimal Lorentz-invariance violating QED model.
%.....................................................
\section{Lee-Wick theory}
%....................................................
In 1969 Lee and Wick\cite{LW} proposed a modified 
QED with the advantage of being finite 
but leading to an indefinite metric in Hilbert space. The origin of the 
negative metric is an extra field introduced by hand, which
may be seen to arise from a higher-order operator as well.\cite{patricio}
Several issues regarding stability and unitarity
were solved using what is now called
the Lee-Wick prescription. 

The point of departure from usual quantum theory is the definition of 
the inner product. In an indefinite metric theory the inner product of
 two states $\vert \phi \rangle$ and $\vert \psi \rangle$
is defined by
$\langle \phi \vert \psi\rangle=\phi^*_i \eta_{ij}\psi_j$
where the metric $\eta_{ij}$ can take negative values. In this way one
has negative norm states in the theory.
In particular, the eigenstates of the self-adjoint hamiltonian operator
can be states with positive norm and real eigenvalues 
 or states with zero norm and complex eigenvalues.\cite{boulware}
In this way the Hilbert space contains states with positive 
norm which oscillate in time and zero norm states which grow or decay.
 The Lee and Wick prescription consists of excising the growing 
or decaying modes from the asymptotic Hilbert space 
and to modify the Feynman diagrams
diagram by diagram
to allow for stability and unitarity of the $S$ matrix.
%..........................................
\section{The QED model}
%.........................................
Our starting point is the Myer and Pospelov lagrangian\cite{MP}
\begin{equation}
\mathcal L=\bar \psi(i {\sla{D}}-m)\, \psi+{ g \bar \psi \sla{n }  
 (n\cdot \partial)^2\psi }-\frac{1}{4}F_{\mu \nu}F^{\mu \nu}    \,,
\end{equation}
with $n$ a privileged four-vector and $g$ a small parameter. 
We choose $n=(1,0,0,0)$ which yields the dispersion relation 
$(p_0-g p_0^2)^2-E^2=0$
with $E = \sqrt{\vec p^2+m^2}$. 
The hamiltonian is
 \begin{eqnarray}\label{Ham}
 H&=& \sum_{s}  \int \frac{d^3 p}{(2\pi)^3} \left( \omega_1  
 a_p^{s\dag}   a_p^{s}  +    
 \omega_2 b_p^{s}   b_p^{s\dag} 
  - W_1  c_p^{s\dag}   c_p^{s}  
 - W_2  d_p^{s}   d_p^{s\dag}   \right) \,,
 \end{eqnarray}
where the four solutions are given by
\begin{eqnarray}
\omega_1=\frac{ 1-\sqrt{1-4gE}}{2g}, \qquad \omega_2=\frac{ 1-\sqrt{1+4g E  }}{2g}, \nonumber 
\\
W_1=\frac{ 1+\sqrt{1-4gE  }}{2g} , \qquad W_2=\frac{ 1+\sqrt{1+4g E }}{2g}.
\end{eqnarray}
The first two solutions $\omega_1,\omega_2$ correspond to modifications to the usual solutions
$E,-E$, respectively, and the next two $W_1,W_2$ correspond to Lee-Wick modes
associated to a negative metric, 
as seen from the hamiltonian \refeq{Ham}.
%...............................................................
\section{Stability and unitarity}
%..............................................................
Our goal is to verify the optical theorem which is basically 
the equality of
the sum over final states of the amplitude 
with the imaginary part of the loop diagram.
We follow the Lee-Wick prescription in order to prove the equality of
 the matrix elements
in the  
scattering process $e^+(k_1)+e^{-}(k_2)\to e^+(k)+e^{-}(k^{\prime})$.
Some central points to
satisfy the perturbative constraint are: (i) the sum over physical states 
in the amplitude diagram must be carried only over positive 
metric states, (ii) in the loop diagram a suitable prescription for the path $C$ in needed 
to avoid the poles and to compute the residues, (iii) the previous prescription has to
reproduce well the usual case in a limit and have the exact discontinuities in the physical sheet in
order to produce the correct imaginary part.
Finally, by comparing both sides one is able to prove the unitarity constraint
 for the considered one-loop scattering process. 

\section*{Acknowledgments}

This work has been partially supported by 
the Chilean research project Fondecyt Regular No.\ 1140781 
and by the group of \emph{F\'{\i}sica de Altas Energ\'{\i}as}
of the Universidad del B{\'i}o-B{\'i}o, Chile.

\end{document}